\documentstyle[12pt]{article}

\def\presentation{      \voffset -.55in
			\hoffset -.19in
			\oddsidemargin 0in
			\evensidemargin 0in
			\marginparwidth .75in
			\marginparsep 7pt
			\topmargin 0in
			\headheight 12pt
			\headsep .25in
			\footheight 18pt
			\footskip .35in
			\textheight 9.5in
			\textwidth 6.5in
			\columnsep 10pt
			\columnseprule 0pt
			\parskip 0pt
			\parsep 0pt
		}
\presentation
\def\Alg{{\cal A}}
\def\ow{\overline{w}}
\let\a=\alpha
\let\b=\beta
\let\la=\lambda
\def\T{\bf T}
\def\Bbb#1{{\bf#1}}

\begin{document}
\bibliographystyle{perso}
\begin{titlepage}
\null \vskip -0.6cm
\vskip 1.4truecm
\begin{center}
		ON THE SYMMETRIES OF INTEGRABILITY
\vskip 1truecm
M. Bellon, J-M. Maillard, C. Viallet
\end{center}
\vfill
\noindent {\bf Abstract.}

We show that the Yang-Baxter equations for two dimensional
models admit as a group of symmetry the infinite discrete group
$A_2^{(1)}$.  The existence of this symmetry explains the presence of a
spectral parameter in the solutions of the equations.  We show that
similarly,
for three-dimensional vertex
models and the associated tetrahedron equations, there also exists an
infinite discrete group of symmetry.  Although generalizing
naturally the previous one, it is a much bigger hyperbolic Coxeter
group.  We indicate how this symmetry can help to resolve the
Yang-Baxter equations and their higher-dimensional generalizations and
initiate the study of three-dimensional vertex models.  These
symmetries are
naturally represented as birational projective transformations.  They
may
preserve non trivial algebraic varieties, and lead to proper
parametrizations of the models, be they integrable or not. We mention
the
 relation   existing between spin models and the   Bose-Messner
 algebras of
 algebraic combinatorics. Our results also yield the generalization
of the condition $q^n=1$ so often mentioned in the theory of quantum
groups,
when no $q$ parameter is available.
\par
\vfill
\vskip 1cm
\centerline{Talk given at the COMO meeting ``Advanced quantum field
theory and Critical Phenomena'' (1991)}
\centerline{ and the LUMINY ``Colloque Verdier sur les syst\`emes
int\'egrables'' (1991)}
 \centerline{work supported by CNRS} \vskip .5truecm
\hrule\vskip .5truecm
\begin{center}
\obeylines
Postal address: %
Laboratoire de Physique Th\'eorique et des Hautes Energies
Universit\'e Paris VI, Tour 16, $1^{\rm er}$ \'etage, bo\^\i te 126.
4 Place Jussieu/ F--75252 PARIS Cedex 05
 \end{center}
\end{titlepage}

\section{Introduction}

The results presented here appear in a series of papers by M.~Bellon,
J-M.~Maillard and C-M.~Viallet~\cite{bmv1,bmv1b,bmv2,bmv2b,prl,prl2}.

The Yang-Baxter equations, which appeared twenty years ago\footnote{
In fact, fifty years ago, Lars Onsager was totally aware of the key
role
played by the star-triangle relation in solving the two-dimensional
Ising model, but he preferred to give an algebraic solution emphasizing
Clifford algebras~\cite{On71,On44,SteMi72,BaEn78,Ka74}.
}, have acquired a predominant role in the theory of integrable
two-dimensional models in statistical mechanics~\cite{Ba80,Ba81} and
field
theory (quantum or classical).  They have actually outpassed the
borders of physics and have become fashionable in some parts of the
mathematics literature.  They in particular support the construction of
quantum groups~\cite{Dr86,BaVi89}.

The Yang-Baxter equations~\cite{Ba81} and their higher dimensional
generalizations are now considered as the defining relations of
integrability.  They are  the  ``Deus ex machina'' in a number of
domains of Mathematics and Physics (Knot Theory~\cite{Jo90}, Quantum
Inverse
Scattering~\cite{Fa82}, S-Matrix Factorization, Exactly Solvable Models
in
Statistical Mechanics, Bethe Ansatz~\cite{Ga83}, Quantum
Groups~\cite{Fa90,BaVi89}, Chromatic
Polynomials~\cite{MaRa83} and more awaited deformation theories).  The
appeal of
these equations  comes from their ability to {\em give
 global results from local ones}.  For instance, they are a  sufficient
and, to some extent,  necessary~\cite{LoMa86} condition for the
commutation of families of transfer matrices of arbitrary size and even
of corner transfer matrices.
{}From the point of view of topology, one may  understand these relations
by considering them as the generators of a large set of {\em discrete
deformations of the lattice}.  This point of view underlies most
studies
in knot theory~\cite{Jo90} and statistical mechanics (${\bf
Z}$-invariance
\cite{Ba78,YaPe87}).

We want to analyze the Yang-Baxter equations and  their higher
dimensional
generalizations~\cite{Ba83,Za81,JaMa82b,Ba86} without prejudice about
what
should be a solution, that is to say proceed by {\em necessary}
conditions,
and have as an input the form of the matrix of Boltzmann weights.

We will exhibit an infinite discrete group of transformations acting on
the Yang-Baxter equations or their higher
dimensional generalizations (tetrahedron, hyper-simplicial equations).

These transformations  act as an automorphy group of various quantities
of
interest in Statistical Mechanics (partition function,\dots), and are
of
great help for calculations, even outside the domain of integrability
(critical manifolds, phase dia\-gram,\dots)~\cite{HaMaOiVe87}.

We show here is that {\em they form a group of symmetries of the
equations defining integrability}.  They consequently appear as a
group
of automorphisms of the Yang-Baxter or tetrahedron equations.  We will
denote this group
${\cal A}ut$.

The existence  of ${\cal A}ut$  drastically constrains the varieties
where solutions may be found.
In the general case, it has {\em infinite} orbits and gives {\em severe
constraints} on the algebraic varieties which parametrize the possible
solutions (genus zero or one curves, algebraic varieties which are not
of the general type~\cite{Ma86}).  In the non-generic case, when ${\cal
A}ut$  has finite order orbits, the algebraic varieties can be of
general type, but {\em the very finiteness condition allows for their
determination}~\cite{bmv1}.

In the framework of infinite group representations, it is crucial to
recognize the {\em essential difference} between what these symmetry
groups
are for the Yang-Baxter equations and what they are for  the higher
dimensional tetrahedron and hyper-simplicial relations: the number of
involutions generating our groups increases from 2 to $2^{d-1}$ when
passing from two-dimensional to $d$-dimensional models and the group
jumps from the semi-direct product  ${\bf Z}\times {\bf Z}_2$ to a
much larger group, i.e., a {\em group with an
exponential growth} with the length of the word.

The existence of  ${\cal A}ut$ as a   symmetry of  the Yang-Baxter
equations has
the following consequence: we may say that solving the Yang-Baxter
equation is equivalent to solving all its images by  ${\cal A}ut$.
These images {\em  generically tend to proliferate}, simply because
${\cal A}ut$ is infinite.  Considering that the equations form an
{\em overdetermined set}, it is easy to believe that the  total set of
equations is ``less overdetermined'' when the orbits of ${\cal A}ut$
are of finite order.  One can therefore  imagine  that the {\em best
candidates} for the integrability varieties are {\em precisely} the
ones where
the symmetry group possesses {\em  finite orbits}: the solutions of
Au-Yang et al.~\cite{AuCoPeTaYa87,CoPeTaSa87,BaPeAu88} seem to confirm
this point of view~\cite{HaMa88,AvMaTaVi90}.

A contrario, if  one gets hold of an apparently isolated  solution, the
action of  ${\cal A}ut$ will multiply it until building up, in
experimentally not so rare cases,  a continuous family of solutions
from the original one.  This is the solution to the so-called
baxterization problem~\cite{bmv2b}.

We first show that the simplest  example of Yang-Baxter relation which
is the star-triangle relation \cite{Ba81} has an {\em infinite discrete
group} of symmetries generated by three  involutions.  These
involutions
are deeply linked with the so-called {\em inversion relations}
\cite{St79,Ba82,JaMa82,Ma84}.

which
two-dimensional
model.

This analysis  can be extended to the ``generalized star-triangle
relation''
for Interaction aRound the Face models without any major
difficulties~\cite{Ba80,MaGa84}.

\section{The star-triangle relations}

\subsection{The setting}

We consider  a spin model with nearest neighbour interactions on square
lattice.  The
spins $\sigma_i$ can take $q$ values.  The Boltzmann weight for an
oriented bond $\langle i j \rangle$ will be denoted hereafter by
$w(\sigma_i,\sigma_j)$.  The weights $w(\sigma_i,\sigma_j)$ can be seen
as the entries of a $q \times q$ matrix.  In the following we will
introduce a  pictorial representation of the star-triangle relation.
An
arrow is  associated to the oriented bond $\langle i j \rangle$.
The arrow from $i$ to $j$ indicates that the argument of the Boltzmann
weight $w$ is $(\sigma_i,\sigma_j)$ rather than $(\sigma_j,\sigma_i)$.
This arrow is  relevant only for the so-called chiral
models~\cite{AuCoPeTaYa87}, that
is to say that the $q\times q$ matrix describing $w$ is not symmetric.
An interesting class of $q\times q$ matrices has been extensively
investigated in the last few years
\cite{AuCoPeTaYa87,BaPeAu88,CoPeTaSa87}:  the general cyclic matrices.
It is important to note that we {\em do not} restrict ourselves to this
particular class of matrices.

The finite algebras we are lead to consider contain in particular
Bose-Messner algebras, i.e.  stuctures occurring in graph theory, and
more
precisely algebraic combinatorics~\cite{BaIt84}. The detailed
relationship
will not be explained here. The reader could compare paragraph 5
of~\cite{bmv1} and the definition of Bose-Messner algebra of
association
schemes in~\cite{BrCoNe89} ,page 44, and~\cite{Ja91}.

Let us give the following
non cyclic nor symmetric $6\times 6$ matrix as  another illustrative
example:
\begin{equation}
  \pmatrix {    x& y& z& y& z& z \cr
		z& x& y& z& y& z  \cr
		y& z& x& z& z& y  \cr
		y& z& z& x& z& y \cr
		z& y& z& y& x& z \cr
		z& z& y& z& y& x \cr
	}    \label{bmv}
\end{equation}

\subsection{The relations}

We introduce the star-triangle equations both analytically\footnote{
Since the $w_i$ and $\ow_i$ are homogeneous variables, there will
always be a
global multiplicative factor $\lambda$ floating around in the
star-triangle
equations.
} and pictorially:
\begin{equation}\label{st1}
	\sum_\sigma w_1(\sigma_1,\sigma) \cdot w_2(\sigma,\sigma_2)
	\cdot
		w_3(\sigma,\sigma_3)
	= \lambda\;  \overline w_1(\sigma_2,\sigma_3) \cdot
		\ow_2(\sigma_1,\sigma_3) \cdot
		\ow_3(\sigma_1,\sigma_2).
\end{equation}
\[
\setlength{\unitlength}{0.0075in}%
\begin{picture}(268,158)( 61,625)
\thicklines
\label{st1p}
\put( 78,700){\vector( 4, 3){ 88}}
\put( 78,700){\vector( 4,-3){ 88}}
\put(296,700){\vector( 1, 2){ 33}}
\put(296,700){\vector( 1,-2){ 33}}
\put(166,766){\vector( 0,-1){132}}
\put(230,700){\vector( 1, 0){ 66}}
\put(152,700){\makebox(0,0)[lb]{\raisebox{0pt}[0pt][0pt]{$\bar 1$}}}
\put(110,649){\makebox(0,0)[lb]{\raisebox{0pt}[0pt][0pt]{$\bar 2$}}}
\put(114,739){\makebox(0,0)[lb]{\raisebox{0pt}[0pt][0pt]{$\bar 3$}}}
\put(329,625){\makebox(0,0)[lb]{\raisebox{0pt}[0pt][0pt]{\tenrm (3)}}}
\put(329,774){\makebox(0,0)[lb]{\raisebox{0pt}[0pt][0pt]{\tenrm (2)}}}
\put(214,703){\makebox(0,0)[lb]{\raisebox{3pt}[0pt][0pt]{\tenrm (1)}}}
\put( 61,703){\makebox(0,0)[lb]{\raisebox{3pt}[0pt][0pt]{\tenrm (1)}}}
\put(164,774){\makebox(0,0)[lb]{\raisebox{0pt}[0pt][0pt]{\tenrm (2)}}}
\put(168,628){\makebox(0,0)[lb]{\raisebox{0pt}[0pt][0pt]{\tenrm (3)}}}
\put(259,706){\makebox(0,0)[lb]{\raisebox{0pt}[0pt][0pt]{$1$}}}
\put(300,736){\makebox(0,0)[lb]{\raisebox{0pt}[0pt][0pt]{$2$}}}
\put(300,662){\makebox(0,0)[lb]{\raisebox{0pt}[0pt][0pt]{$3$}}}
\put(185,700){\makebox(0,0)[lb]{\raisebox{0pt}[0pt][0pt]{\svtnrm =}}}
\put(510,700){\makebox(0,0)[lb]{\raisebox{0pt}[0pt][0pt]{\tenrm
(st1.1)}}}
\end{picture}
\]
One should note that satisfying  equation (\ref{st1}) {\em together
with}
the relation (st1.2) obtained by reversing all arrows, is a {\em
sufficient}
condition for the commutation of the  diagonal transfer matrices of
{\em
arbitrary size}  $M$  with periodic boundary conditions
${\T}_M(w_2,\ow_2)$
and ${\T}_M(\ow_3,w_3)$:
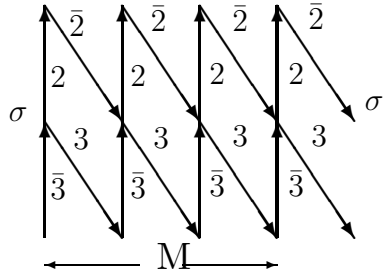
\[
\setlength{\unitlength}{0.0075in}%
\begin{picture}(248,185)(56,617)
\thinlines
\put(177,620){\vector( 1, 0){ 66}}
\put(147,620){\vector(-1, 0){ 66}}
\thicklines
\put(243,801){\vector( 2,-3){ 54}}
\put( 81,720){\vector( 0, 1){ 81}}
\put( 81,801){\vector( 2,-3){ 54}}
\put( 81,720){\vector( 2,-3){ 54}}
\put( 81,639){\vector( 0, 1){ 81}}
\put(135,639){\vector( 0, 1){ 81}}
\put(135,720){\vector( 2,-3){ 54}}
\put(135,801){\vector( 2,-3){ 54}}
\put(135,720){\vector( 0, 1){ 81}}
\put(189,720){\vector( 0, 1){ 81}}
\put(189,801){\vector( 2,-3){ 54}}
\put(189,720){\vector( 2,-3){ 54}}
\put(189,639){\vector( 0, 1){ 81}}
\put(243,639){\vector( 0, 1){ 81}}
\put(243,720){\vector( 2,-3){ 54}}
\put(243,720){\vector( 0, 1){ 81}}
\put(159,617){\makebox(0,0)[lb]{\raisebox{0pt}[0pt][0pt]{\frtnrm M}}}
\put( 56,720){\makebox(0,0)[lb]{\raisebox{0pt}[0pt][0pt]{\elvrm
$\sigma$}}}
\put(304,728){\makebox(0,0)[lb]{\raisebox{0pt}[0pt][0pt]{\elvrm
$\sigma$}}}
\put(265,785){\makebox(0,0)[lb]{\raisebox{0pt}[0pt][0pt]{\elvrm
$\bar2$}}}
\put(212,783){\makebox(0,0)[lb]{\raisebox{0pt}[0pt][0pt]{\elvrm
$\bar2$}}}
\put(155,783){\makebox(0,0)[lb]{\raisebox{0pt}[0pt][0pt]{\elvrm
$\bar2$}}}
\put( 98,780){\makebox(0,0)[lb]{\raisebox{0pt}[0pt][0pt]{\elvrm
$\bar2$}}}
\put( 85,744){\makebox(0,0)[lb]{\raisebox{0pt}[0pt][0pt]{\elvrm $2$}}}
\put( 85,666){\makebox(0,0)[lb]{\raisebox{0pt}[0pt][0pt]{\elvrm
$\bar3$}}}
\put(101,701){\makebox(0,0)[lb]{\raisebox{0pt}[0pt][0pt]{\elvrm $3$}}}
\put(157,702){\makebox(0,0)[lb]{\raisebox{0pt}[0pt][0pt]{\elvrm $3$}}}
\put(141,667){\makebox(0,0)[lb]{\raisebox{0pt}[0pt][0pt]{\elvrm
$\bar3$}}}
\put(141,745){\makebox(0,0)[lb]{\raisebox{0pt}[0pt][0pt]{\elvrm $2$}}}
\put(196,746){\makebox(0,0)[lb]{\raisebox{0pt}[0pt][0pt]{\elvrm $2$}}}
\put(196,668){\makebox(0,0)[lb]{\raisebox{0pt}[0pt][0pt]{\elvrm
$\bar3$}}}
\put(212,703){\makebox(0,0)[lb]{\raisebox{0pt}[0pt][0pt]{\elvrm $3$}}}
\put(267,703){\makebox(0,0)[lb]{\raisebox{0pt}[0pt][0pt]{\elvrm $3$}}}
\put(251,668){\makebox(0,0)[lb]{\raisebox{0pt}[0pt][0pt]{\elvrm
$\bar3$}}}
\put(251,746){\makebox(0,0)[lb]{\raisebox{0pt}[0pt][0pt]{\elvrm $2$}}}
\end{picture}
\]

Note that for cyclic matrices (\cite{AuCoPeTaYa87,BaPeAu88,CoPeTaSa87})
the star-triangle relations (st1.1) and (st1.2) give the same equations
since
one exchanges (st1.2) and  (st1.1) by spin reversal.

One could obviously imagine many other choices for the arrows on the
six
bonds, however only three of them lead to the commutation of diagonal
transfer matrices.  We therefore have three systems of equations to
study.
For example, if the Boltzmann weights are given by  the $6\times 6$
matrix (\ref{bmv}), these three systems of equations are respectively
made of
20 different equations or 35 or 36.

\section{The Yang-Baxter relation for vertex models.}

We shall not get here into the arcanes of this relation, which appears
in
the theory of integrable models~\cite{BaVi89}, the theory of
factorizable
$S$-matrix in two-dimensional field theory, the quantum inverse
scattering
method~\cite{Fa82}, knot theory and has been given a canonical meaning
in
terms of Hopf algebras~\cite{Ab77} (quantum
groups~\cite{Dr86,BaVi89,ob90,Ve87,Ji85}) and the list is far from
exhaustive.  We just want to fix some notations for later use.

We consider a vertex model on  a two-dimensional square lattice.  To
each
bond is associated a variable with $q$ possible states and a Boltzmann
weight $w(i,j,k,l)$ is assigned to each vertex
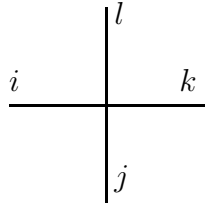
\[
\setlength{\unitlength}{0.0085in}%
\begin{picture}(120,131)(20,700)
\thicklines
\put( 80,820){\line( 0,-1){120}}
\put( 20,760){\line( 1, 0){120}}
\put(125,769){\makebox(0,0)[lb]{\raisebox{0pt}[0pt][0pt]{\elvrm $k$}}}
\put( 85,710){\makebox(0,0)[lb]{\raisebox{0pt}[0pt][0pt]{\elvrm $j$}}}
\put( 85,810){\makebox(0,0)[lb]{\raisebox{0pt}[0pt][0pt]{\elvrm $l$}}}
\put( 20,769){\makebox(0,0)[lb]{\raisebox{0pt}[0pt][0pt]{\elvrm $i$}}}
\end{picture}
\]

In order to write the Yang-Baxter relation, the $q^4$ homogeneous
weights $w(i,j,k,l)$ are first arranged in a $q^2\times q^2$ matrix
$R$:
\begin{equation}
	R^{ij}_{kl}=w(i,j,k,l).         \label{ass}
\end{equation}
The Yang-Baxter relation is a trilinear relation between three matrices
$R(1,2)$, $R(2,3)$ and $R(1,3)$:
\begin{equation}   \label{yb}
	\sum_{\alpha_1,\alpha_2,\alpha_3} R^{i_1i_2}_{\a_1\a_2} (1,2)
		R^{\a_1i_3}_{j_1\a_3} (1,3) R^{\a_2\a_3}_{j_2j_3} (2,3)
     =\sum_{\b_1,\b_2,\b_3} R^{i_2i_3}_{\b_2\b_3} (2,3)
	  R^{i_1\b_3}_{\b_1j_3} (1,3) R^{\b_1\b_2}_{j_1j_2} (1,2).
\end{equation}
The assignation (\ref{ass}) is arbitrary and we may specify it by
complementing the vertex with an arrow and attributing numbers to the
lines
$$
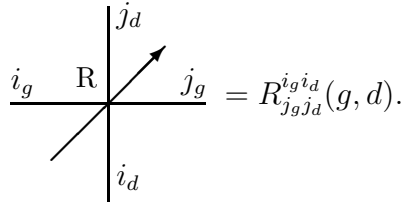
\raisebox{-0.51in}{
\setlength{\unitlength}{0.0085in}%
\begin{picture}(120,131)(20,700)
\thicklines
\put( 45,725){\vector( 1, 1){ 70}}
\put( 80,820){\line( 0,-1){120}}
\put( 20,760){\line( 1, 0){120}}
\put(125,769){\makebox(0,0)[lb]{\raisebox{0pt}[0pt][0pt]{\elvrm
$j_g$}}}
\put( 85,710){\makebox(0,0)[lb]{\raisebox{0pt}[0pt][0pt]{\elvrm
$i_d$}}}
\put( 85,810){\makebox(0,0)[lb]{\raisebox{0pt}[0pt][0pt]{\elvrm
$j_d$}}}
\put( 20,769){\makebox(0,0)[lb]{\raisebox{0pt}[0pt][0pt]{\elvrm
$i_g$}}}
\put( 60,770){\makebox(0,0)[lb]{\raisebox{0pt}[0pt][0pt]{\elvrm R}}}
\end{picture}
} = R^{i_gi_d}_{j_gj_d} (g,d).
$$
With these rules  relation (\ref{yb}) has the following graphical
representation
\begin{equation} \label{picyb}
\raisebox{-0.51in}{
\setlength{\unitlength}{0.0085in}%
\begin{picture}(425,236)(25,565)
\thicklines
\put( 40,720){\line( 3,-2){180}}
\put( 40,660){\line( 3, 2){180}}
\put(170,800){\line( 0,-1){220}}
\put( 55,690){\vector( 1, 0){ 60}}
\put(150,620){\vector( 3, 2){ 45}}
\put(155,715){\vector( 1, 2){ 30}}
\put(305,605){\vector( 1, 2){ 30}}
\put(300,735){\vector( 3, 2){ 45}}
\put(375,690){\vector( 1, 0){ 60}}
\put(320,800){\line( 0,-1){220}}
\put(450,660){\line(-3, 2){180}}
\put(450,720){\line(-3,-2){180}}
\put(360,640){\makebox(0,0)[lb]{\raisebox{0pt}[0pt][0pt]{\elvrm
$\b_2$}}}
\put(360,730){\makebox(0,0)[lb]{\raisebox{0pt}[0pt][0pt]{\elvrm
$\b_1$}}}
\put(315,690){\makebox(0,0)[rb]{\raisebox{0pt}[0pt][0pt]{\elvrm
$\b_3$}}}
\put(280,615){\makebox(0,0)[b]{\raisebox{0pt}[0pt][0pt]{\elvrm $i_2$}}}
\put(280,755){\makebox(0,0)[b]{\raisebox{0pt}[0pt][0pt]{\elvrm $i_1$}}}
\put(325,595){\makebox(0,0)[lb]{\raisebox{0pt}[0pt][0pt]{\elvrm
$i_3$}}}
\put(325,790){\makebox(0,0)[lb]{\raisebox{0pt}[0pt][0pt]{\elvrm
$j_3$}}}
\put(200,750){\makebox(0,0)[lb]{\raisebox{0pt}[0pt][0pt]{\elvrm
$j_2$}}}
\put(200,620){\makebox(0,0)[lb]{\raisebox{0pt}[0pt][0pt]{\elvrm
$j_1$}}}
\put(440,650){\makebox(0,0)[rb]{\raisebox{0pt}[0pt][0pt]{\elvrm
$j_1$}}}
\put(440,720){\makebox(0,0)[rb]{\raisebox{0pt}[0pt][0pt]{\elvrm
$j_2$}}}
\put(165,780){\makebox(0,0)[rb]{\raisebox{0pt}[0pt][0pt]{\elvrm
$j_3$}}}
\put(165,595){\makebox(0,0)[rb]{\raisebox{0pt}[0pt][0pt]{\elvrm
$i_3$}}}
\put(175,685){\makebox(0,0)[lb]{\raisebox{0pt}[0pt][0pt]{\elvrm
$\a_3$}}}
\put(120,725){\makebox(0,0)[b]{\raisebox{0pt}[0pt][0pt]{\elvrm
$\a_2$}}}
\put(120,650){\makebox(0,0)[b]{\raisebox{0pt}[0pt][0pt]{\elvrm
$\a_1$}}}
\put( 50,650){\makebox(0,0)[b]{\raisebox{0pt}[0pt][0pt]{\elvrm $i_2$}}}
\put( 50,720){\makebox(0,0)[b]{\raisebox{0pt}[0pt][0pt]{\elvrm $i_1$}}}
\put(245,685){\makebox(0,0)[b]{\raisebox{0pt}[0pt][0pt]{\twlrm =}}}
\put( 25,720){\makebox(0,0)[lb]{\raisebox{0pt}[0pt][0pt]{\twlrm 1}}}
\put( 25,660){\makebox(0,0)[lb]{\raisebox{0pt}[0pt][0pt]{\twlrm 2}}}
\put(155,565){\makebox(0,0)[lb]{\raisebox{0pt}[0pt][0pt]{\twlrm 3}}}
\put(305,570){\makebox(0,0)[lb]{\raisebox{0pt}[0pt][0pt]{\twlrm 3}}}
\put(260,585){\makebox(0,0)[lb]{\raisebox{0pt}[0pt][0pt]{\twlrm 2}}}
\put(260,785){\makebox(0,0)[lb]{\raisebox{0pt}[0pt][0pt]{\twlrm 1}}}
\end{picture}
}
\end{equation}
The lines carry indices 1,2,3.

Some especially interesting solutions depend on a continuous parameter
called the ``spectral parameter''.  The presence of this parameter is
fundamental for many applications in physics, as for example the Bethe
Ansatz method~\cite{LiWu72,Ka74,Fa82,Ga83}.  One of the main issues in
the full
resolution of (\ref{yb}) is precisely to describe what is this
parameter and
the algebraic variety on which it lives, although its presence may
obscure
the algebraic structures underlying the Yang-Baxter equation ({\em the
discovery of quantum groups was allowed by forgetting this
parameter}~\cite{Ji85,Dr86,Wor87,BaVi89}).  The problem of building up
continuous families of solutions from an isolated one, known as the
{\em
baxterization}~\cite{Jo90}, is made straightforward by our study.
Indeed
our results {\em explain the presence of the spectral parameter} in the
solution of the equation (see also~\cite{bmv1}).

\section{The  symmetry group of the star-triangle
relation}
\subsection{The inversion relation}

Two distinct inverses act on the matrix of nearest neighbour spin
interactions: the
matrix inverse $I$ and the dyadic (element by element) inverse $J$.  We
write down the inversion relations both analytically and pictorially:
\begin{eqnarray}\label{i1}
	 \sum_\sigma w(\sigma_i,\sigma) \cdot I(w)(\sigma,\sigma_j) &=&
		\mu \; \delta_{\sigma_i \sigma_j},    \\
	w(\sigma_i,\sigma_j) \cdot J(w)(\sigma_i,\sigma_j) &=& 1.
	\label{i2}
\end{eqnarray}
where $ \delta_{\sigma_i \sigma_j}$ denotes the usual Kronecker delta.

\bigskip
\setlength{\unitlength}{0.0075in}%
\begin{picture}(315,40)(58,740)
\thicklines
\put(235,760){\circle*{7}}
\put(160,760){\circle*{7}}
\put( 80,760){\circle*{7}}
\put(320,760){\circle*{7}}
\put(200,760){\line( 1, 0){ 40}}
\put(120,760){\vector( 1, 0){ 80}}
\put( 80,760){\vector( 1, 0){ 40}}
\put(320,740){\makebox(0,0)[b]{\raisebox{0pt}[0pt][0pt]{\tenrm
$\sigma_i=\sigma_j$}}}
\put(235,740){\makebox(0,0)[b]{\raisebox{0pt}[0pt][0pt]{\tenrm
$\sigma_j$}}}
\put( 80,740){\makebox(0,0)[b]{\raisebox{0pt}[0pt][0pt]{\tenrm
$\sigma_i$}}}
\put(155,740){\makebox(0,0)[b]{\raisebox{0pt}[0pt][0pt]{\tenrm
$\sigma$}}}
\put(180,770){\makebox(0,0)[lb]{\raisebox{0pt}[0pt][0pt]{\tenrm
$I(w)$}}}
\put(105,770){\makebox(0,0)[lb]{\raisebox{0pt}[0pt][0pt]{\tenrm $w$}}}
\put(270,760){\makebox(0,0)[lb]{\raisebox{0pt}[0pt][0pt]{\frtnrm =}}}
\end{picture}
 \raisebox{2.5ex}{ and \ } \raisebox{-1.4ex}{
\setlength{\unitlength}{0.0065in}%
\begin{picture}(422,94)(126,675)
\thicklines
\put(540,720){\circle*{8}}
\put(402,720){\circle*{8}}
\put(290,720){\circle*{8}}
\put(130,720){\circle*{8}}
\put(210,720){\oval(160,60)}
\put(210,750){\vector( 1, 0){  5}}
\put(210,690){\vector( 1, 0){  5}}
\put(539,703){\makebox(0,0)[rb]{\raisebox{0pt}[0pt][0pt]{\elvrm
$\sigma_j$}}}
\put(402,701){\makebox(0,0)[lb]{\raisebox{0pt}[0pt][0pt]{\elvrm
$\sigma_i$}}}
\put(139,712){\makebox(0,0)[lb]{\raisebox{0pt}[0pt][0pt]{\elvrm
$\sigma_i$}}}
\put(281,713){\makebox(0,0)[rb]{\raisebox{0pt}[0pt][0pt]{\elvrm
$\sigma_j$}}}
\put(345,720){\makebox(0,0)[b]{\raisebox{0pt}[0pt][0pt]{\frtnrm =}}}
\put(210,760){\makebox(0,0)[b]{\raisebox{0pt}[0pt][0pt]{\tenrm $w$}}}
\put(205,660){\makebox(0,0)[t]{\raisebox{0pt}[0pt][0pt]{\tenrm
$J(w)$}}}
\end{picture}
	}
\bigskip

The two involutions $I$ and $J$ generate an {\em infinite discrete
group $\Gamma$}
(Coxeter group) isomorphic to the infinite dihedral group ${\bf Z}_2
\times {\bf Z}$.  The ${\bf Z}$ part of $\Gamma$ is generated by $IJ$.
In the parameter space of the model, that is to say some projective
space
${\bf C}{\bf P}_{n-1}$ ($n$ homogeneous parameters), $I$ and $J$ are
birational involutions.  They give a {\em non-linear representation of
this
Coxeter group by an infinite set of birational
transformations}~\cite{bmv1}.
It  may happen that the action of $\Gamma$ on specific subvarieties
{\em
yields a finite orbit}.  This means that the representation of $\Gamma$
identifies with the $p$-dihedral group ${\bf Z}_2 \times {\bf Z}_p$.

\subsection{The symmetries of the star-triangle relations}
\label{sostr}

The two inversions $I$ and $J$ act on the star-triangle relation.  Let
us give a pictorial representation of this action, starting from
(st1.1) as an example:\par
\bigskip
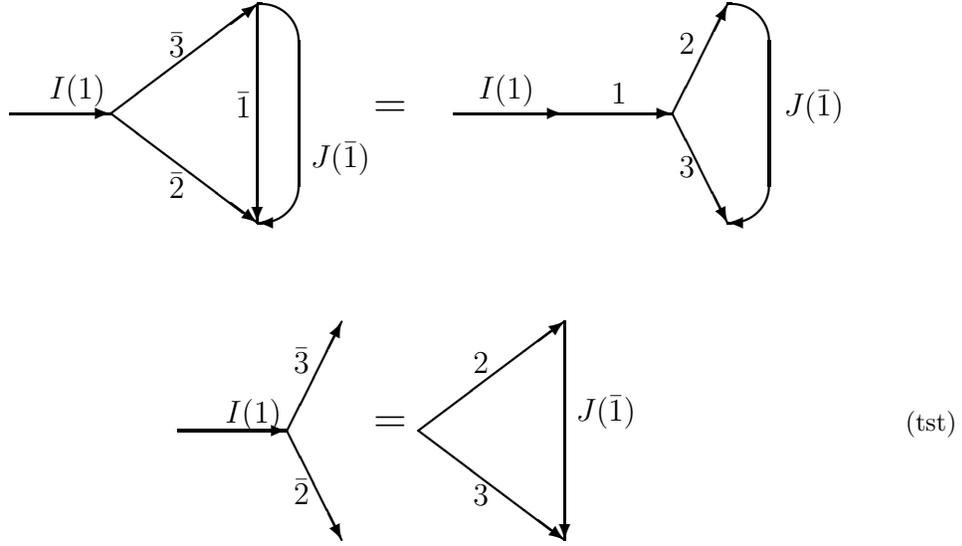
\null\hfill
\setlength{\unitlength}{0.0087in}%
\begin{picture}(466,384)(41,324)
\thicklines
\put(473,575){\vector(-1, 0){0}}
\put(473,641){\oval( 50,132)[br]}
\put(473,641){\oval( 50,132)[tr]}
\put(190,575){\vector(-1, 0){0}}
\put(190,641){\oval( 50,132)[br]}
\put(190,641){\oval( 50,132)[tr]}
\put(308,641){\vector( 1, 0){ 66}}
\put( 41,641){\vector( 1, 0){ 61}}
\put(190,707){\vector( 0,-1){132}}
\put(102,641){\vector( 4, 3){ 88}}
\put(102,641){\vector( 4,-3){ 88}}
\put(374,641){\vector( 1, 0){ 66}}
\put(440,641){\vector( 1, 2){ 33}}
\put(440,641){\vector( 1,-2){ 33}}
\put(287,450){\vector( 4, 3){ 88}}
\put(287,450){\vector( 4,-3){ 88}}
\put(375,516){\vector( 0,-1){132}}
\put(208,450){\vector( 1, 2){ 33}}
\put(208,450){\vector( 1,-2){ 33}}
\put(142,450){\vector( 1, 0){ 66}}
\put( 65,650){\makebox(0,0)[lb]{\raisebox{0pt}[0pt][0pt]{$I(1)$}}}
\put(323,650){\makebox(0,0)[lb]{\raisebox{0pt}[0pt][0pt]{$I(1)$}}}
\put(507,640){\makebox(0,0)[lb]{\raisebox{0pt}[0pt][0pt]{$ J(\bar1)$}}}
\put(222,609){\makebox(0,0)[lb]{\raisebox{0pt}[0pt][0pt]{$ J(\bar1)$}}}
\put(260,449){\makebox(0,0)[lb]{\raisebox{0pt}[0pt][0pt]{\svtnrm =}}}
\put(260,639){\makebox(0,0)[lb]{\raisebox{0pt}[0pt][0pt]{\svtnrm =}}}
\put(137,675){\makebox(0,0)[lb]{\raisebox{0pt}[0pt][0pt]{$\bar 3$}}}
\put(137,590){\makebox(0,0)[lb]{\raisebox{0pt}[0pt][0pt]{$\bar 2$}}}
\put(177,639){\makebox(0,0)[lb]{\raisebox{0pt}[0pt][0pt]{$\bar 1$}}}
\put(444,603){\makebox(0,0)[lb]{\raisebox{0pt}[0pt][0pt]{$3$}}}
\put(444,677){\makebox(0,0)[lb]{\raisebox{0pt}[0pt][0pt]{$2$}}}
\put(403,647){\makebox(0,0)[lb]{\raisebox{0pt}[0pt][0pt]{$1$}}}
\put(382,456){\makebox(0,0)[lb]{\raisebox{0pt}[0pt][0pt]{$ J(\bar1)$}}}
\put(320,405){\makebox(0,0)[lb]{\raisebox{0pt}[0pt][0pt]{$3$}}}
\put(320,485){\makebox(0,0)[lb]{\raisebox{0pt}[0pt][0pt]{$2$}}}
\put(171,455){\makebox(0,0)[lb]{\raisebox{0pt}[0pt][0pt]{$I(1)$}}}
\put(212,485){\makebox(0,0)[lb]{\raisebox{0pt}[0pt][0pt]{$\bar 3$}}}
\put(212,406){\makebox(0,0)[lb]{\raisebox{0pt}[0pt][0pt]{$\bar 2$}}}
\put(580,450){\makebox(0,0)[lb]{\raisebox{0pt}[0pt][0pt]{\tenrm
(tst)}}}
\end{picture}
\hfill\null
\bigskip
\par
The transformed equation  reads:
\begin{equation}
\label{mst2}
	\lambda\; \sum_{\sigma_1} I(w_1)(\tau,\sigma_1)
		\cdot \ow_2(\sigma_1,\sigma_3)\cdot
		\ow_3(\sigma_1,\sigma_2)=
	w_2(\tau,\sigma_2) \cdot w_3(\tau,\sigma_3)\cdot
		J(\ow_1)(\sigma_2,\sigma_3).
\end{equation}

We get an action on the space of solutions of the star-triangle
relation.

If $(w_1,w_2,w_3,\ow_1,\ow_2,\ow_3)$ is a solution of
eq(\ref{st1}) (see  picture (st1.1) for the specific arrangement of
arrows), then $(I(w_1),\ow_3,\ow_2,J(\ow_1),w_3,w_2)$ is also a
solution of eq(\ref{st1}),  at the price of a permitted redefinition of
$\lambda$.  In this transformation, the weights $w_1$ and $\ow_1$ play
a special role.

At this point, it is better to formalize this action by introducing
some
notations.  We may choose as a reference star-triangle relation ${\cal
ST}$,
the  symmetric configuration:
\par\bigskip
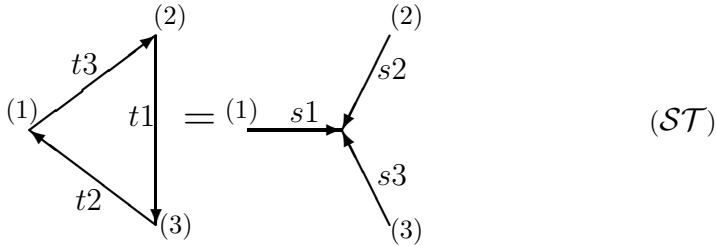
\setlength{\unitlength}{0.0075in}%
\begin{picture}(268,158)(-31,625)
\thicklines
\put( 78,700){\vector( 4, 3){ 88}}
\put(166,634){\vector(-4, 3){ 88}}
\put(329,766){\vector(-1,-2){ 33}}
\put(329,634){\vector(-1, 2){ 33}}
\put(166,766){\vector( 0,-1){132}}
\put(230,700){\vector( 1, 0){ 66}}
\put(147,703){\makebox(0,0)[lb]{\raisebox{0pt}[0pt][0pt]{$t1$}}}
\put(110,644){\makebox(0,0)[lb]{\raisebox{0pt}[0pt][0pt]{$t2$}}}
\put(107,739){\makebox(0,0)[lb]{\raisebox{0pt}[0pt][0pt]{$t3$}}}
\put(329,625){\makebox(0,0)[lb]{\raisebox{0pt}[0pt][0pt]{\tenrm (3)}}}
\put(329,774){\makebox(0,0)[lb]{\raisebox{0pt}[0pt][0pt]{\tenrm (2)}}}
\put(214,703){\makebox(0,0)[lb]{\raisebox{3pt}[0pt][0pt]{\tenrm (1)}}}
\put( 61,703){\makebox(0,0)[lb]{\raisebox{3pt}[0pt][0pt]{\tenrm (1)}}}
\put(164,774){\makebox(0,0)[lb]{\raisebox{0pt}[0pt][0pt]{\tenrm (2)}}}
\put(168,628){\makebox(0,0)[lb]{\raisebox{0pt}[0pt][0pt]{\tenrm (3)}}}
\put(259,706){\makebox(0,0)[lb]{\raisebox{0pt}[0pt][0pt]{$s1$}}}
\put(320,736){\makebox(0,0)[lb]{\raisebox{0pt}[0pt][0pt]{$s2$}}}
\put(320,662){\makebox(0,0)[lb]{\raisebox{0pt}[0pt][0pt]{$s3$}}}
\put(185,700){\makebox(0,0)[lb]{\raisebox{0pt}[0pt][0pt]{\svtnrm =}}}
\put(510,700){\makebox(0,0)[lb]{\raisebox{0pt}[0pt][0pt]{\tenrm (${\cal
ST}$)}}}
\end{picture}
\par\bigskip

Any configuration may be obtained by reversing some arrows and
permuting some  bonds.  With evident notations, we will denote by
$R_{\rm s1},R_{\rm s2},R_{\rm s3},R_{\rm t1},R_{\rm t2},R_{\rm t3}$ the
reversals of arrows, and by $P_{\rm si,sj},P_{\rm si,tj},P_{\rm ti,tj}$
the permutations of bonds.  Moreover $I$ and $J$ act on the bonds as
$I_{\rm s1},I_{\rm s2},\dots$
The action of $I$ and $J$ described above (where 1 was playing a
special role) identifies with the action of
\begin{equation}
	{\cal K}_1= R_{\rm s2} R_{\rm t3} I_{\rm s1} J_{\rm t1} P_{\rm
	s2,t3}
		P_{\rm s3,t2}.
\end{equation}
It is easy to check that ${\cal K}_1$ is an involution.

We may  construct two similar involutions ${\cal K}_2$ and
${\cal K}_3$, obtained by cyclic permutation of the indices 1,~2,~3.
The involutions  ${\cal K}_i (i=1,2,3)$ verify the defining relations
of the {\it Weyl group of an affine algebra of type}
$A_2^{(1)}$~\cite{Ka85}:
\begin{eqnarray}
	({\cal K}_1{\cal K}_2)^3 = ({\cal K}_2{\cal K}_3)^3 =
		({\cal K}_3{\cal K}_1)^3 = 1.
\end{eqnarray}
We denote ${\cal A}ut$ the group generated by the three
involutions  ${\cal K}_i \;(i=1,2,3)$.

\section{The symmetry group of  the Yang-Baxter equation.}
\subsection{The inversion relations.}\label{notations}

The $R$-matrix appears naturally as a representation of an element of
the
tensor product $\Alg\otimes\Alg$ of some algebra $\Alg$ with itself.
This algebra is a nice Hopf algebra in the context of quantum groups.
We shall not dwell on this here but recall some simple operations on
$R$.

In $\Alg\otimes\Alg$ we have a product inherited from the product in
$\Alg$:
\begin{equation}
	(a\otimes b) (c\otimes d) = ac \otimes bd.
\end{equation}
$R$ is an invertible element of $\Alg\otimes\Alg$ for this product and
we
shall denote by $I(R)$ the inverse for this product:
\begin{equation}
	R\cdot I(R) = I(R) \cdot R = 1\otimes1.
\end{equation}
In terms of the representative matrix this reads:
\begin{equation}
	\sum_{\a,\b} R^{ij}_{\a\b}
	\;I(R)^{\a\b}_{uv}=\delta^i_u\;\delta^j_v=
		\sum_{\a,\b} I(R)^{ij}_{\a\b} \;R^{\a\b}_{uv} .
\end{equation}
This is nothing else but the so-called {\em inversion relation} for
{\em  vertex}
models~\cite{St79,Ba82,MaGa84,Ma83,Ma86}.
On  $\Alg\otimes\Alg$ we have a permutation operator $\sigma$:
\begin{eqnarray}
	\sigma( a \otimes b ) &=& b \otimes a,  \\
	( \sigma R ) ^{ij}_{uv} &=& R^{ji}_{vu}, \quad\mbox{ for the
	matrix $R$.}
\end{eqnarray}
Note that the representation of $\sigma$ is just the conjugation by the
permutation matrix $P$:
\begin{eqnarray}
	P^{ij}_{kl} &=& \delta_{il} \delta_{jk},  \\
	\sigma R &=& PRP.
\end{eqnarray}

In the language of matrices we have a notion of transposition.  Let us
define
partial transpositions $t_g$ and $t_d$ by:
\begin{eqnarray}
	(t_g R)^{ij}_{uv} &=& R^{uj}_{iv},              \\
	(t_d R)^{ij}_{uv} &=& R_{uj}^{iv},
\end{eqnarray}
and the full transposition
\begin{equation}
	t = t_g t_d = t_d t_g.
\end{equation}
We shall in the sequel use another inversion $J$ defined by:
\begin{equation}
	J = t_g I t_d= t_d I t_g,       \label{col}
\end{equation}
or equivalently:
\begin{equation}
	\sum_{\a,\b} R^{\a u}_{v \b} \;J(R)^{\a i}_{j\b}= \delta^i_u
	\;\delta^j_v=
		\sum_{\a,\b} J(R)_{\a j}^{i\b} \;R_{\a v}^{u \b}
\end{equation}
These operators verify straightforwardly:
\begin{eqnarray}
	I^2 &=& J^2 =1, \quad It=tI, \quad Jt=tJ, \nonumber\\
	\sigma^2 &=& t^2 =1, \quad \sigma I = I \sigma, \quad \sigma J
	= J \sigma,
\nonumber\\
	(\sigma t_g)^2 &=& (\sigma t_d)^2 =t, \quad \sigma t_g \sigma
	t_d = 1.
\end{eqnarray}
Each of these operations has a graphical representation.
For the inversion $I$ or more precisely for $\sigma I$ it is:
\begin{equation}
\setlength{\unitlength}{0.0085in}%
\begin{picture}(367,132)(55,621)
\thicklines
\put(281,666){\line( 1, 1){ 79}}
\put(320,625){\line( 1, 1){ 79}}
\put(130,695){\vector( 1, 1){ 22}}
\put( 90,655){\vector( 1, 1){ 20}}
\put(100,625){\line( 0, 1){ 80}}
\put(100,705){\line( 1, 0){ 80}}
\put( 60,665){\line( 1, 0){ 80}}
\put(140,665){\line( 0, 1){ 80}}
\put(413,700){\makebox(0,0)[b]{\raisebox{0pt}[0pt][0pt]{\elvrm $v$}}}
\put(367,741){\makebox(0,0)[lb]{\raisebox{0pt}[0pt][0pt]{\elvrm $u$}}}
\put(275,668){\makebox(0,0)[b]{\raisebox{0pt}[0pt][0pt]{\elvrm $i$}}}
\put(315,621){\makebox(0,0)[rb]{\raisebox{0pt}[0pt][0pt]{\elvrm $j$}}}
\put(240,675){\makebox(0,0)[b]{\raisebox{0pt}[0pt][0pt]{\frtnrm =}}}
\put( 62,669){\makebox(0,0)[b]{\raisebox{0pt}[0pt][0pt]{\elvrm $i$}}}
\put(179,710){\makebox(0,0)[b]{\raisebox{0pt}[0pt][0pt]{\elvrm $v$}}}
\put(144,691){\makebox(0,0)[lb]{\raisebox{0pt}[0pt][0pt]{\elvrm $\sigma
I(R)$}}}
\put(104,649){\makebox(0,0)[lb]{\raisebox{0pt}[0pt][0pt]{\elvrm $R$}}}
\put( 93,622){\makebox(0,0)[rb]{\raisebox{0pt}[0pt][0pt]{\elvrm $j$}}}
\put(145,742){\makebox(0,0)[lb]{\raisebox{0pt}[0pt][0pt]{\elvrm $u$}}}
\put(145,665){\makebox(0,0)[b]{\raisebox{0pt}[0pt][0pt]{\elvrm $\a$}}}
\put(100,710){\makebox(0,0)[b]{\raisebox{0pt}[0pt][0pt]{\elvrm $\b$}}}
\end{picture}
   \label{figI}
\end{equation}
the inversion $J$ reads:
\begin{equation}
\setlength{\unitlength}{0.0085in}%
\begin{picture}(366,136)(54,621)
\thicklines
\put(153,689){\vector(-3, 4){ 24}}
\put(113,649){\vector(-3, 4){ 24}}
\put(320,625){\line( 1, 1){ 79}}
\put(281,666){\line( 1, 1){ 79}}
\put(100,625){\line( 0, 1){ 80}}
\put(100,705){\line( 1, 0){ 80}}
\put( 60,665){\line( 1, 0){ 80}}
\put(140,665){\line( 0, 1){ 80}}
\put(240,675){\makebox(0,0)[b]{\raisebox{0pt}[0pt][0pt]{\frtnrm =}}}
\put(315,621){\makebox(0,0)[rb]{\raisebox{0pt}[0pt][0pt]{\elvrm $u$}}}
\put(413,696){\makebox(0,0)[b]{\raisebox{0pt}[0pt][0pt]{\elvrm $i$}}}
\put(343,744){\makebox(0,0)[lb]{\raisebox{0pt}[0pt][0pt]{\elvrm $j$}}}
\put(279,674){\makebox(0,0)[b]{\raisebox{0pt}[0pt][0pt]{\elvrm $v$}}}
\put( 63,671){\makebox(0,0)[b]{\raisebox{0pt}[0pt][0pt]{\elvrm $v$}}}
\put(179,710){\makebox(0,0)[b]{\raisebox{0pt}[0pt][0pt]{\elvrm $i$}}}
\put(144,691){\makebox(0,0)[lb]{\raisebox{0pt}[0pt][0pt]{\elvrm $
J(R)$}}}
\put(104,649){\makebox(0,0)[lb]{\raisebox{0pt}[0pt][0pt]{\elvrm $\sigma
R$}}}
\put( 98,623){\makebox(0,0)[rb]{\raisebox{0pt}[0pt][0pt]{\elvrm $u$}}}
\put(145,742){\makebox(0,0)[lb]{\raisebox{0pt}[0pt][0pt]{\elvrm $j$}}}
\put(145,665){\makebox(0,0)[lb]{\raisebox{0pt}[0pt][0pt]{\elvrm $\a$}}}
\put(100,710){\makebox(0,0)[b]{\raisebox{0pt}[0pt][0pt]{\elvrm $\b$}}}
\end{picture}
 \label{figJ}
\end{equation}
and the transposition reads:
\[
\setlength{\unitlength}{0.0085in}%
\begin{picture}(285,121)(20,700)
\thicklines
\put(270,735){\vector(-1, 1){ 50}}
\put(245,820){\line( 0,-1){120}}
\put(185,760){\line( 1, 0){120}}
\put( 20,760){\line( 1, 0){120}}
\put( 80,820){\line( 0,-1){120}}
\put( 55,735){\vector( 1, 1){ 50}}
\put(160,760){\makebox(0,0)[lb]{\raisebox{0pt}[0pt][0pt]{\elvrm =}}}
\put(250,765){\makebox(0,0)[lb]{\raisebox{0pt}[0pt][0pt]{\elvrm $A$}}}
\put( 85,745){\makebox(0,0)[lb]{\raisebox{0pt}[0pt][0pt]{\elvrm $\sigma
t_dA$}}}
\put(290,765){\makebox(0,0)[lb]{\raisebox{0pt}[0pt][0pt]{\elvrm $k$}}}
\put(250,710){\makebox(0,0)[lb]{\raisebox{0pt}[0pt][0pt]{\elvrm $j$}}}
\put(250,810){\makebox(0,0)[lb]{\raisebox{0pt}[0pt][0pt]{\elvrm $l$}}}
\put(185,765){\makebox(0,0)[lb]{\raisebox{0pt}[0pt][0pt]{\elvrm $i$}}}
\put( 20,765){\makebox(0,0)[lb]{\raisebox{0pt}[0pt][0pt]{\elvrm $i$}}}
\put( 85,810){\makebox(0,0)[lb]{\raisebox{0pt}[0pt][0pt]{\elvrm $l$}}}
\put( 85,710){\makebox(0,0)[lb]{\raisebox{0pt}[0pt][0pt]{\elvrm $j$}}}
\put(125,765){\makebox(0,0)[lb]{\raisebox{0pt}[0pt][0pt]{\elvrm $k$}}}
\end{picture}
\]

Note that the two inversions $I$ and $J$ do not commute.  They generate
an
infinite discrete group $\Gamma$, the infinite dihedral group,
isomorphic to
the semi-direct product ${\bf Z} \times {\bf Z}_2$.  This group is
represented on the matrix elements by { \em birational
transformations}~\cite{bmv1,bmv3,bmv5} acting on the projective space
of the
entries of the matrix $R$.  Remark that for the {\em vertex models},
the
birational transformations associated to the two involutions $I$ and
$J$ are
naturally related by collineations (see (\ref{col}): this should be
compared
with the situation for nearest neighbour interaction spin
models~\cite{bmv1,Sh77}.

\subsection{The symmetries of the Yang-Baxter equations.}
\label{sybe}

At the price of the redefinitions:
\begin{eqnarray}
	A &=& t R(2,3),         \\
	B &=& \sigma t_d R(1,3),        \\
	C &=& R(1,2),
\end{eqnarray}
we may picture the Yang-Baxter relation in a
more symmetric way:
\begin{equation} \label{ybref}
\raisebox{-1cm}{
\setlength{\unitlength}{0.0085in}%
\begin{picture}(425,235)(25,565)
\thicklines
\put(334,661){\vector(-1,-2){ 30}}
\put(186,778){\vector(-1,-2){ 30}}
\put(334,719){\vector(-1, 2){ 30}}
\put(185,604){\vector(-1, 2){ 30}}
\put( 40,720){\line( 3,-2){180}}
\put( 40,660){\line( 3, 2){180}}
\put(170,800){\line( 0,-1){220}}
\put( 55,690){\vector( 1, 0){ 60}}
\put(375,690){\vector( 1, 0){ 60}}
\put(320,800){\line( 0,-1){220}}
\put(450,660){\line(-3, 2){180}}
\put(450,720){\line(-3,-2){180}}
\put(405,697){\makebox(0,0)[b]{\raisebox{0pt}[0pt][0pt]{$C$}}}
\put(329,751){\makebox(0,0)[b]{\raisebox{0pt}[0pt][0pt]{$ B$}}}
\put(181,637){\makebox(0,0)[b]{\raisebox{0pt}[0pt][0pt]{$ B$}}}
\put(310,635){\makebox(0,0)[b]{\raisebox{0pt}[0pt][0pt]{$ A$}}}
\put(180,735){\makebox(0,0)[b]{\raisebox{0pt}[0pt][0pt]{$ A$}}}
\put( 84,699){\makebox(0,0)[b]{\raisebox{0pt}[0pt][0pt]{$ C$}}}
\put(245,685){\makebox(0,0)[b]{\raisebox{0pt}[0pt][0pt]{\twlrm =}}}
\put( 25,720){\makebox(0,0)[lb]{\raisebox{0pt}[0pt][0pt]{\twlrm 1}}}
\put( 25,660){\makebox(0,0)[lb]{\raisebox{0pt}[0pt][0pt]{\twlrm 2}}}
\put(155,565){\makebox(0,0)[lb]{\raisebox{0pt}[0pt][0pt]{\twlrm 3}}}
\put(305,570){\makebox(0,0)[lb]{\raisebox{0pt}[0pt][0pt]{\twlrm 3}}}
\put(260,585){\makebox(0,0)[lb]{\raisebox{0pt}[0pt][0pt]{\twlrm 2}}}
\put(260,785){\makebox(0,0)[lb]{\raisebox{0pt}[0pt][0pt]{\twlrm 1}}}
\end{picture}
}
\end{equation}

We may bracket (\ref{ybref}) with \raisebox{-0.2in}{
\setlength{\unitlength}{0.0085in}%
\begin{picture}(105,70)(25,660)
\thicklines
\put( 40,720){\line( 3,-2){ 90}}
\put( 40,660){\line( 3, 2){ 90}}
\put( 55,690){\vector( 1, 0){ 60}}
\put( 85,710){\makebox(0,0)[b]{\raisebox{0pt}[0pt][0pt]{\elvrm $\tilde
C$}}}
\put( 25,720){\makebox(0,0)[lb]{\raisebox{0pt}[0pt][0pt]{\twlrm 1}}}
\put( 25,660){\makebox(0,0)[lb]{\raisebox{0pt}[0pt][0pt]{\twlrm 2}}}
\end{picture}
}, where
$\tilde C= \sigma I (C)$.
We get
\begin{equation}
\raisebox{-1cm}{
\setlength{\unitlength}{0.0085in}%
\begin{picture}(544,138)(46,641)
\thicklines
\put(535,710){\vector( 1, 0){ 40}}
\put(240,710){\vector( 1, 0){ 40}}
\put(291,664){\line(-2, 3){ 60}}
\put(231,664){\line( 2, 3){ 60}}
\put(111,665){\line( 2, 3){ 60}}
\put(171,665){\line(-2, 3){ 60}}
\put(120,711){\vector( 1, 0){ 40}}
\put( 60,710){\vector( 1, 0){ 40}}
\put(111,664){\line(-2, 3){ 60}}
\put( 51,664){\line( 2, 3){ 60}}
\put(172,754){\line( 1, 0){ 58}}
\put(172,663){\line( 1, 0){ 59}}
\put(200,779){\line( 0,-1){138}}
\put(213,648){\vector(-3, 4){ 24}}
\put(213,768){\vector(-1,-1){ 25}}
\put(425,675){\vector( 1,-1){ 25}}
\put(425,740){\vector( 3, 4){ 24}}
\put(436,779){\line( 0,-1){138}}
\put(464,663){\line(-1, 0){ 59}}
\put(464,754){\line(-1, 0){ 58}}
\put(585,664){\line(-2, 3){ 60}}
\put(525,664){\line( 2, 3){ 60}}
\put(475,710){\vector( 1, 0){ 40}}
\put(465,665){\line( 2, 3){ 60}}
\put(525,665){\line(-2, 3){ 60}}
\put(405,664){\line(-2, 3){ 60}}
\put(345,664){\line( 2, 3){ 60}}
\put(355,710){\vector( 1, 0){ 40}}
\put(316,707){\makebox(0,0)[b]{\raisebox{0pt}[0pt][0pt]{\elvrm =}}}
\put(261,722){\makebox(0,0)[b]{\raisebox{0pt}[0pt][0pt]{\elvrm $\tilde
C$}}}
\put(141,723){\makebox(0,0)[b]{\raisebox{0pt}[0pt][0pt]{\elvrm $C$}}}
\put( 81,722){\makebox(0,0)[b]{\raisebox{0pt}[0pt][0pt]{\elvrm $\tilde
C$}}}
\put(191,760){\makebox(0,0)[b]{\raisebox{0pt}[0pt][0pt]{\elvrm $A$}}}
\put(192,648){\makebox(0,0)[b]{\raisebox{0pt}[0pt][0pt]{\elvrm $B$}}}
\put(423,759){\makebox(0,0)[b]{\raisebox{0pt}[0pt][0pt]{\elvrm $B$}}}
\put(426,648){\makebox(0,0)[b]{\raisebox{0pt}[0pt][0pt]{\elvrm $A$}}}
\put(555,722){\makebox(0,0)[b]{\raisebox{0pt}[0pt][0pt]{\elvrm $\tilde
C$}}}
\put(495,723){\makebox(0,0)[b]{\raisebox{0pt}[0pt][0pt]{\elvrm $C$}}}
\put(375,722){\makebox(0,0)[b]{\raisebox{0pt}[0pt][0pt]{\elvrm $\tilde
C$}}}
\end{picture}
}
\label{bra}
\end{equation}
that is to say
\begin{equation}
\raisebox{-1cm}{
\setlength{\unitlength}{0.0085in}%
\begin{picture}(425,235)(25,565)
\thicklines
\put(334,661){\vector(-1,-2){ 30}}
\put(186,778){\vector(-1,-2){ 30}}
\put(334,719){\vector(-1, 2){ 30}}
\put(185,604){\vector(-1, 2){ 30}}
\put( 40,720){\line( 3,-2){180}}
\put( 40,660){\line( 3, 2){180}}
\put(170,800){\line( 0,-1){220}}
\put( 55,690){\vector( 1, 0){ 60}}
\put(375,690){\vector( 1, 0){ 60}}
\put(320,800){\line( 0,-1){220}}
\put(450,660){\line(-3, 2){180}}
\put(450,720){\line(-3,-2){180}}
\put(340,625){\makebox(0,0)[b]{\raisebox{0pt}[0pt][0pt]{\frtnrm $t_g
A$}}}
\put(335,752){\makebox(0,0)[b]{\raisebox{0pt}[0pt][0pt]{\frtnrm $t_d
B$}}}
\put(405,710){\makebox(0,0)[b]{\raisebox{0pt}[0pt][0pt]{\elvrm
$tI(C)$}}}
\put( 85,710){\makebox(0,0)[b]{\raisebox{0pt}[0pt][0pt]{\elvrm
$tI(C)$}}}
\put(190,637){\makebox(0,0)[b]{\raisebox{0pt}[0pt][0pt]{\frtnrm $t_d
B$}}}
\put(190,735){\makebox(0,0)[b]{\raisebox{0pt}[0pt][0pt]{\frtnrm $t_g
A$}}}
\put(245,685){\makebox(0,0)[b]{\raisebox{0pt}[0pt][0pt]{\twlrm =}}}
\put( 25,720){\makebox(0,0)[lb]{\raisebox{0pt}[0pt][0pt]{\twlrm 1}}}
\put( 25,660){\makebox(0,0)[lb]{\raisebox{0pt}[0pt][0pt]{\twlrm 2}}}
\put(155,570){\makebox(0,0)[lb]{\raisebox{0pt}[0pt][0pt]{\twlrm 3}}}
\put(305,570){\makebox(0,0)[lb]{\raisebox{0pt}[0pt][0pt]{\twlrm 3}}}
\put(260,585){\makebox(0,0)[lb]{\raisebox{0pt}[0pt][0pt]{\twlrm 2}}}
\put(260,785){\makebox(0,0)[lb]{\raisebox{0pt}[0pt][0pt]{\twlrm 1}}}
\end{picture}
}
\end{equation}
This relation is nothing but (\ref{ybref}) after the redefinitions
\begin{eqnarray}
	A &\rightarrow& t_g A,  \nonumber\\
	B &\rightarrow& t_d B,  \nonumber\\
	C &\rightarrow& tI\>C.  \label{K3}
\end{eqnarray}
We may denote by $K_3$ the operation (\ref{K3}).  We have two other
similar operations $K_1$ and $K_2$
\[
     \begin{array}{rl}
		K_1:    &A\rightarrow   tI\>A   \\
			&B\rightarrow   t_g B   \\
			&C\rightarrow   t_d C
     \end{array}    \quad,\quad
     \begin{array}{rl}
		K_2:    &A \rightarrow  t_d A   \\
			&B \rightarrow  tI\>B   \\
			&C \rightarrow  t_g C
     \end{array}    \quad.
\]
The discrete group ${\cal A}ut$ generated by the $K_i$'s ($i=1,2,3$) is
a symmetry group of the Yang-Baxter equations.  These generators $K_i$
($i=1,2,3$) are involutions.
The $K_i$'s satisfy the relation $(K_1K_2K_3)^2 = 1$.  Actually, the
operation
$K_1K_2K_3$ is just the inversion $I$ on $R$.
Among the elements of the discrete group generated by the $K_i$'s we
have in
particular:
\begin{eqnarray}
(K_1K_2)^2:\quad        A &\rightarrow& It_g It_g A = t IJ A,   \\
	B &\rightarrow& t_d I t_d I B = t JI B, \\
	C &\rightarrow& C.
\end{eqnarray}
Since $IJ$ is of infinite order, we have generated an {\em infinite
discrete
group} of symmetries.  This is exactly the phenomenon that we described
in section \ref{sostr} for the star-triangle equations.

Under this form it is not so evident to find the actual structure of
the
group.  Let us introduce $K_A$, $K_B$ and $K_C$, which are simply
related to the $K_i$'s by the transposition of two vertices:
\[
	\begin{array}{rl}
		K_A:    &A      \rightarrow     \sigma tI A     \\
			&B      \rightarrow     t_g \sigma C    \\
			&C      \rightarrow     \sigma t_g B
	\end{array}     \quad,\quad
	\begin{array}{rl}
		K_B:    &A      \rightarrow     \sigma t_g C
		\\
			&B      \rightarrow     \sigma tI B
			\\
			&C      \rightarrow     t_g \sigma A
	\end{array}     \quad,\quad
	\begin{array}{rl}
		K_C:    &A      \rightarrow     t_g \sigma B
		\\
			&B      \rightarrow     \sigma t_g A    \\
			&C      \rightarrow     \sigma tI C
	\end{array}     \quad.
\]
It is easily verified that:
\begin{equation}
	K_A^2 = K_B^2 = K_C^2 =1,
\end{equation}
and
\begin{equation}
	(K_AK_B)^3 = (K_BK_C)^3 = (K_CK_A)^3 =1,
\end{equation}
with no other relations.  We recover the affine Coxeter group
$A_2^{(1)}$ we
already encountered in section \ref{sostr}.

\bigskip

We have here a very powerful instrument:
it defines {\em adequate patterns} for the matrix $R$~\cite{bmv4}.
It permits the so-called {\em baxterization of an isolated solution}
just
acting with $tIJ$.
Indeed if a set of relations among the entries of $R$ are preserved by
$IJ$ (or at least by $tIJ$), they will stay for every transforms of the
initial Yang-Baxter relation.
We shall illustrate in  section \ref{bobm} the baxterization on
the Baxter eight-vertex model~\cite{Ba72,Ba78} and show
in section \ref{bosln} how to introduce a spectral parameter for the
solutions of
the Yang-Baxter equations associated to $sl(n)$ algebras.

\section{The tetrahedron equations and their symmetries.}

This equation is a generalization of the Yang-Baxter equation to three
dimensional vertex models~\cite{Za81,Ba83,Ba86}.
We give a pictorial representation of the three-dimensional vertex by
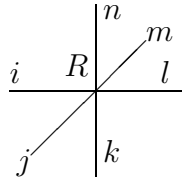
\[
\setlength{\unitlength}{0.0075in}%
\begin{picture}(120,120)(20,700)
\thinlines
\put( 35,715){\line( 1, 1){ 80}}
\put( 80,820){\line( 0,-1){120}}
\put( 20,760){\line( 1, 0){120}}
\put(115,795){\makebox(0,0)[lb]{\raisebox{0pt}[0pt][0pt]{\elvrm $m$}}}
\put( 25,705){\makebox(0,0)[lb]{\raisebox{0pt}[0pt][0pt]{\elvrm $j$}}}
\put(125,765){\makebox(0,0)[lb]{\raisebox{0pt}[0pt][0pt]{\elvrm $l$}}}
\put( 85,710){\makebox(0,0)[lb]{\raisebox{0pt}[0pt][0pt]{\elvrm $k$}}}
\put( 85,810){\makebox(0,0)[lb]{\raisebox{0pt}[0pt][0pt]{\elvrm $n$}}}
\put( 20,765){\makebox(0,0)[lb]{\raisebox{0pt}[0pt][0pt]{\elvrm $i$}}}
\put( 75,770){\makebox(0,0)[rb]{\raisebox{0pt}[0pt][0pt]{\elvrm $R$}}}
\end{picture}
\]
The Boltzmann weights of the vertex are denoted $w(i,j,k,l,m,n)$ and
may be arranged in a matrix of entries
\begin{equation}
	R^{ijk}_{lmn} = w (i,j,k,l,m,n).
\label{3R}
\end{equation}
The tetrahedron equation has a pictorial representation:
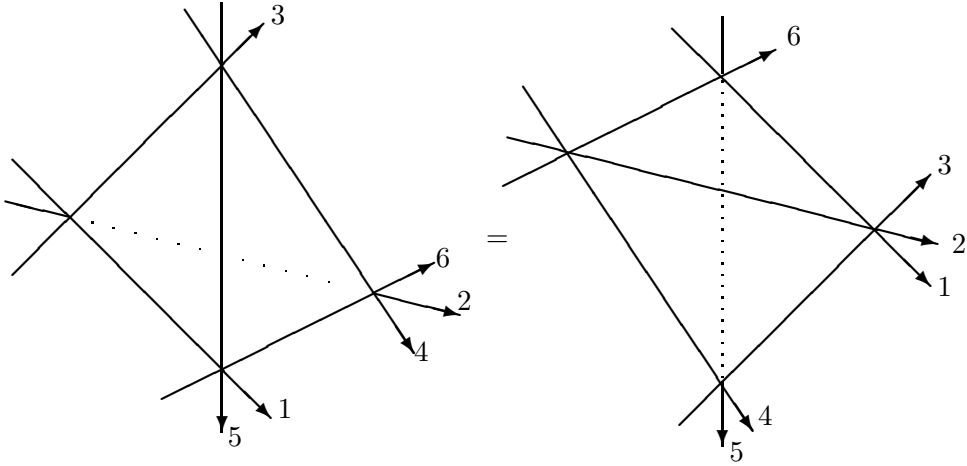
\[
\setlength{\unitlength}{0.0075in}%
\begin{picture}(660,316)(15,470)
\thicklines
\put( 20,600){\vector( 1, 1){175}}
\put( 20,680){\vector( 1,-1){180}}
\put(166,790){\vector( 0,-1){300}}
\put(124,513){\vector( 2, 1){190}}
\put(140,785){\vector( 2,-3){160}}
\put( 60,640){\line(-4, 1){ 45}}
\put(272,587){\vector( 4,-1){ 60}}
\put(365,696){\vector( 4,-1){300}}
\put(376,731){\vector( 2,-3){160}}
\put(362,662){\vector( 2, 1){190}}
\put(515,525){\vector( 0,-1){ 45}}
\put(515,740){\line( 0, 1){ 40}}
\put(480,772){\vector( 1,-1){180}}
\put(485,495){\vector( 1, 1){175}}
\thinlines
\multiput( 75,636.25)(15,-3.75){12}{\line( 1, 0){ 1}}
\multiput(515,735)(0.00000,-7.92453){27}{\line( 0,-1){  1}}
\put(520,470){\makebox(0,0)[lb]{\raisebox{0pt}[0pt][0pt]{\elvrm 5}}}
\put(540,495){\makebox(0,0)[lb]{\raisebox{0pt}[0pt][0pt]{\elvrm 4}}}
\put(665,585){\makebox(0,0)[lb]{\raisebox{0pt}[0pt][0pt]{\elvrm 1}}}
\put(675,615){\makebox(0,0)[lb]{\raisebox{0pt}[0pt][0pt]{\elvrm 2}}}
\put(665,670){\makebox(0,0)[lb]{\raisebox{0pt}[0pt][0pt]{\elvrm 3}}}
\put(560,760){\makebox(0,0)[lb]{\raisebox{0pt}[0pt][0pt]{\elvrm 6}}}
\put(200,775){\makebox(0,0)[lb]{\raisebox{0pt}[0pt][0pt]{\elvrm 3}}}
\put(315,605){\makebox(0,0)[lb]{\raisebox{0pt}[0pt][0pt]{\elvrm 6}}}
\put(330,575){\makebox(0,0)[lb]{\raisebox{0pt}[0pt][0pt]{\elvrm 2}}}
\put(300,540){\makebox(0,0)[lb]{\raisebox{0pt}[0pt][0pt]{\elvrm 4}}}
\put(205,500){\makebox(0,0)[lb]{\raisebox{0pt}[0pt][0pt]{\elvrm 1}}}
\put(170,480){\makebox(0,0)[lb]{\raisebox{0pt}[0pt][0pt]{\elvrm 5}}}
\put(350,620){\makebox(0,0)[lb]{\raisebox{0pt}[0pt][0pt]{\elvrm =}}}
\end{picture}
\]
The algebraic form is
\begin{equation}
	R_{123} R_{543} R_{516} R_{426} =  R_{426}  R_{516}  R_{543}
	R_{123}.
	\label{tetra}
\end{equation}
We may here again introduce an inverse $I$
\begin{equation}
	\sum_{\a_g,\a_m,\a_d} (IR)_{\a_g\a_m\a_d}^{i_gi_mi_d} \cdot
		R^{\a_g\a_m\a_d}_{j_gj_mj_d} = \delta^{i_g}_{j_g}
			\delta^{i_m}_{j_m}\delta^{i_d}_{j_d}.
			\label{Itetra}
\end{equation}
We also introduce the partial transpositions $t_g$, $t_m$ and $t_d$
with
\begin{equation}
	(t_gR)^{i_gi_mi_d}_{j_gj_mj_d} = R^{j_gi_mi_d}_{i_gj_mj_d},
\end{equation}
and similar definitions for $t_m$ and $t_d$.

We redefine
\begin{equation}
	A       =       R_{123},                \quad
	B       =       t_d R_{543},    \quad
	C       =       t_gt_m R_{516}, \quad
	D       =       t R_{426},
\end{equation}
where $t$ is the full transposition $t_gt_mt_d$.  Equation
(\ref{tetra})
then takes the more symmetric form
\begin{equation}
	\sum_{s_1,\ldots,s_6} A^{i_1i_2i_3}_{s_1s_2s_3}
	B^{i_5i_4j_3}_{s_5s_4s_3} C^{j_5j_1i_6}_{s_5s_1s_6}
	D^{j_4j_2j_6}_{s_4s_2s_6} =  \sum_{r_1,\ldots,r_6}
	D_{i_4i_2i_6}^{r_4r_2r_6} C_{i_5i_1j_6}^{r_5r_1r_6}
	B_{j_5j_4i_3}^{r_5r_4r_3} A_{j_1j_2j_3}^{r_1r_2r_3}.
\label{tetrasym}
\end{equation}
We may multiply the previous equation by $(IA)_{i_1i_2i_3}^{u_1u_2u_3}$
and $(tIA)_{j_1j_2j_3}^{v_1v_2v_3}$ and sum over $(i_1,i_2,i_3)$ and
$(j_1,j_2,j_3)$.  This amounts to a bracketing of the tetrahedron
equations
by two times the same vertex, in a procedure trivially generalizing the
one
for the Yang-Baxter equation (\ref{bra}).  We recover (\ref{tetrasym})
with
$A$, $B$, $C$ and $D$ transformed by
\begin{eqnarray}
	K_1:\;  A &\rightarrow& tIA             \nonumber\\
			B &\rightarrow& t_dB    \nonumber\\
			C &\rightarrow& t_mC    \nonumber\\
			D &\rightarrow& t_mD.
\end{eqnarray}
We have in a similar way the operations
\[
	\begin{array}{rl}
		K_2:    &A      \rightarrow     t_d A   \\
				&B      \rightarrow     tI B
				\\
				&C      \rightarrow     t_g C   \\
				&D      \rightarrow     t_g D
	\end{array}     \quad,\quad
	\begin{array}{rl}
		K_3:    &A      \rightarrow     t_g A   \\
				&B      \rightarrow     t_g B
				\\
				&C      \rightarrow     tI C
				\\
				&D      \rightarrow     t_d D
	\end{array}     \quad,\quad
	\begin{array}{rl}
		K_4:    &A      \rightarrow     t_m A   \\
				&B      \rightarrow     t_m B
				\\
				&C      \rightarrow     t_d C   \\
				&D      \rightarrow     tI D
	\end{array}     \quad.
\]
Each of these four operations is an involution.  They satisfy various
relations, for instance $(K_1K_2K_3K_4)^2=1$.  {\em The $K_i$'s
generate a group ${\cal A}ut_{\bf 3}$ which is a symmetry group of
the tetrahedron equations}.  This group is ``monstrous'' since the
number
of elements of length smaller than $l$ is of exponential growth with
respect to $l$, unlike the case of the affine Coxeter groups (as
$A_2^{(1)}$ for the Yang-Baxter equation) where this number is of
polynomial growth.

The operations playing a role
similar to the one of $I$ and $J$ in the two-dimensional Yang-Baxter
equations are the {\em four} involutions
\begin{equation}
	I,\quad J=t_gIt_mt_d, \quad K=t_mIt_dt_g, \quad L=t_dIt_gt_m.
\end{equation}
We call $\Gamma_3$ the group generated by these four involutions.
$\Gamma_3$ is  also a symmetry group for the three
dimensional vertex model {\em even if}~\cite{JaMa82} the model {\em
does not}
satisfy the tetrahedron equation.

In order to precise the algebraic structure of the group $\Gamma_3$
generated by $I$, $J$, $K$ and~$L$, it is simpler to consider as
generators two of the partial transpositions $t_g$ and $t_d$, $I$ and
the full
transposition $t$.  The third partial transposition can be recovered as
the product $tt_gt_d$ and $t$ commutes with all other generators and so
contributes a mere ${\bf Z}_2$ factor in the group.  We are thus
considering the Coxeter group generated by three involutions $t_g$,
$t_d$ and $I$, with two of them commuting: this is represented by the
following Dynkin diagram
\[
\setlength{\unitlength}{0.0075in}%
\begin{picture}(183,36)(28,777)
\thinlines
\put(200,797){\circle*{6}}
\put(120,797){\circle*{6}}
\put( 40,797){\circle*{6}}
\put( 40,797){\line( 1, 0){160}}
\put(160,803){\makebox(0,0)[b]{\raisebox{0pt}[0pt][0pt]{\elvrm
$\infty$}}}
\put(120,778){\makebox(0,0)[b]{\raisebox{0pt}[0pt][0pt]{\elvrm $I$}}}
\put(200,778){\makebox(0,0)[b]{\raisebox{0pt}[0pt][0pt]{\elvrm $t_d$}}}
\put( 40,778){\makebox(0,0)[b]{\raisebox{0pt}[0pt][0pt]{\elvrm $t_g$}}}
\put( 80,803){\makebox(0,0)[b]{\raisebox{0pt}[0pt][0pt]{\elvrm
$\infty$}}}
\end{picture}
\]
For this group again,
the number of elements of length smaller than $l$ is greater than
$2^{l/2}$.  This is in fact a {\em hyperbolic} Coxeter
group~\cite{Hu90}.

\section{Use of the symmetry group.}

\subsection{The baxterization}

The problem of the baxterization is to introduce a spectral parameter
into an
isolated solution of the Yang-Baxter equations~\cite{Jo90}.  We have
solutions
of this problem by acting with the symmetry group $\Gamma$.

\subsubsection{Baxterization of the Baxter model}
\label{bobm}

Consider the matrix of the symmetric eight vertex model
\begin{equation}
	R=\left( \begin{array}{cccc} a&0&0&d\\ 0&b&c&0\\0&c&b&0\\
	d&0&0&a
			\end{array} \right).
\label{baxter}
\end{equation}
Notice that this form is preserved by $I$ and $J$ and that
$tR=R$.  The action of $I$ is
\begin{eqnarray}
	a &\rightarrow& {a\over a^2-d^2} \qquad
		b \;\rightarrow \;{b\over b^2-c^2}      \\
	c &\rightarrow& {-c\over b^2-c^2} \qquad
		d \;\rightarrow \;{-d\over a^2-d^2}     \\
\noalign{\noindent and the action of $J$ is}
	a &\rightarrow& {a\over a^2-c^2} \qquad
		b \;\rightarrow \;{b\over b^2-d^2}      \\
	c &\rightarrow& {-c\over a^2-c^2} \qquad
		d \;\rightarrow \;{-d\over b^2-d^2}
\end{eqnarray}
We shall look at the solutions of the Yang-Baxter equations for
matrices $R$
of the form~(\ref{baxter}).  The leading idea is that the
parametrization of
the solutions is just the parametrization of the algebraic varieties
preserved
by $tIJ$ in the projective space ${\bf CP}_3$ of the homogenous
parameters
$(a,b,c,d)$.  The remarkable fact is that not only these varieties
exist but can be completely described.  We use the visualization method
we have already used~\cite{bmv1,bmv1b} for spin models, that is to say
just
draw the orbits obtained by numerical iteration and look.

This is best illustrated by figure~1.  This figure shows the orbit of
point (*), which is a matrix of the form~(\ref{baxter}).  It is drawn
by the iteration of $IJ$ acting on the initial point (*).  The
resulting points densify on the elliptic curve given by the
intersection of the two quadrics $\Delta_1 = \mbox{constant}$ and
$\Delta_2
= \mbox{constant}$ (Clebsch's biquadratic), with $\Delta_1$ and
$\Delta_2$ the $\Gamma$ invariants
\begin{eqnarray}
	\Delta_1 &=& {a^2 + b^2 - c^2 - d^2 \over ab+cd },
	\nonumber\\
	\Delta_2 &=& { ab - cd \over ab + cd }.
\label{inv}
\end{eqnarray}

Similar calculations can be performed for a  general 16-vertex model
for
which:
\begin{equation}
	R=\pmatrix{       a_1 &   a_2 &   b_1 &   b_2 \cr
		a_3 &   a_4 &   b_3 &   b_4 \cr
		c_1 &   c_2 &   d_1 &   d_2 \cr
		c_3 &   c_4 &   d_3 &   d_4 \cr }
\label{r16}
\end{equation}

Amazingly the baxterization of the 16-vertex model leads to {\em
curves}. These curves are also {\em
intersection of quadrics} (even in the general case for which their is
no solution for the
Yang-Baxter equations), and lead to a remarkable elliptic
parametrization of the model~\cite{prl2}.

\subsubsection{Baxterization of the  $R$ matrix of $sl_q(n)$}
\label{bosln}

Another example corresponds to the baxterization of solutions
associated to
$sl(n)$ algebras~\cite{Fa90}.  There are special solutions generally
denoted
$R_+$ and $R_-$.  For the simplest four-dimensional representation
of the $sl(2)$ case, we have
\begin{equation}
	R_+ =\left( \begin{array}{cccc} q&0&0&0\\ 0&1&q-q^{-1}&0\\
		0&0&1&0\\  0&0&0&q \end{array} \right).
\label{r+r-}
\end{equation}
and a similar expression for $R_-$~\cite{Fa90}.  Looking for a family
containing both $R_+$ and $R_-$ our baxterization procedure leads to
the
well-known~\cite{Ka74} six-vertex model $R$-matrix $R=\la R_+  + 1/\la
R_-$.

We let as an exercise for the reader to treat the $sl(3)$ case.  In a
forthcoming publication we will show that these ideas can be
generalized to
all the universal $R$-matrices~\cite{BaVi89} for every
representation~\cite{bmvuni}. This group  appears in field theory, in
the analysis
of classical $R$-matrices~\cite{Av91}.

\subsubsection{$q$ root of unity}
One of the most studied cases of quantum group is obtained when the
parameter
$q$ is a root of unity. The structure of the representation theory is
then
extremely rich and differs from the generic one.
\proclaim Proposition: $q^n=1$ is equivalent to: the orbit of $\Gamma$
is finite. \par
This applies {\em even if the parameter $q$ is not defined} (e.g. in
the elliptic
case). See~\cite{bmvuni}.


\subsection{Three dimensional models}

Our strategy for finding solutions  of the tetrahedron equations is to
seek
for patterns of the Boltzmann weights  of the  three dimensional vertex
{\em
compatible with the symmetry group $\Gamma_3$}.  By this we mean that
its
form should be preserved by $\Gamma_3$.

\subsubsection{A first model}

We will therefore consider a simple model where $i$, $j$, $k$, $l$, $m$
and~$n$ take only two values $+1$ and $-1$.  The matrix (\ref{3R}) is
an
$8\times8$ matrix.  We will require that its pattern is invariant under
the
inverse $I$~\cite{bmv4} and the various partial transpositions $t_g$,
$t_m$
and $t_d$.  We aim at having a generalization of the Baxter
eight-vertex
model and we impose the following restrictions:  \begin{eqnarray}
	w(i,j,k,l,m,n) &=& w(-i,-j,-k,-l,-m,-n),
	\label{sr4}\\ w(i,j,k,l,m,n) &=& 0 \qquad\mbox{if
	\ $ijklmn=-1$.}
	\label{kirch} \end{eqnarray} These constraints amount to saying
	that
the $8\times8$ matrix splits into two times the same $4\times4$
matrix.  It
is further possible to impose that this matrix is symmetric since, in
this
case, $t_gR$ (and any other partial transpose) is also symmetric.  Let
us
introduce the following notations for the entries of the $4\times4$
block of
the $R$ matrix \begin{equation}
	\left( \begin{array}{cccc}
				a   & d_1 & d_2 & d_3 \\ d_1 & b_1 &
				c_3 &
				c_2 \\ d_2 & c_3 & b_2 & c_1 \\ d_3 &
				c_2 &
				c_1 & b_3
	\end{array} \right).  \label{R4} \end{equation}
The four rows and columns of this matrix correspond to the states
$(+,+,+)$,
$(+,-,-)$, $(-,+,-)$ and~$(-,-,+)$ for the triplets $(i,j,k)$ or
$(l,m,n)$.
The $R$-matrix can be completed by spin reversal, according to the rule
(\ref{sr4}).  $t_g$  simply exchanges $c_2$ with $d_2$ and $c_3$ with
$d_3$,
$t_m$ and $t_d$ can be similarly defined and $I$ acts as the inversion
of
this $4\times4$ matrix.

For this three dimensional model, the coefficients of the
characteristic
polynomial of the $4\times 4$ matrix (\ref{R4}) give a good hint for
invariants under $\Gamma_3$.  They are
\begin{eqnarray}
	\sigma_1^{(3d)} &=& a + b_1 + b_2 + b_3 , \\
	\sigma_2^{(3d)} &=& a( b_1 + b_2 + b_3 ) +  b_1b_2 + b_2b_3 +
	b_3b_1
		- ( c_1^2+c_2^2+c_3^2+d_1^2+d_2^2+d_3^2), \\
	\ldots. \nonumber
\end{eqnarray}
Since $\sigma_2^{(3d)}$ is invariant by $t_g$, $t_m$ amd $t_d$ and
takes a
simple factor (the inverse of the determinant) under the action of $I$,
the
variety $\sigma_2^{(3d)}=0$ {\em is invariant under $\Gamma_3$}.  Given
the
hugeness of the group $\Gamma_3$, it is already  an astonishing fact to
have
such a covariant expression.  In fact we can exhibit {\em five linearly
independent polynomials} with the same covariance, which give {\em four
invariants}, as follows:
\begin{equation}
	a b_1 + b_2 b_3 - c_1^2 -d_1^2, \qquad
	c_2 d_2 - c_3 d_3,
\end{equation}
and the ones deduced by permutations of 1, 2 and~3.  They form a five
dimensional space of polynomials. Any ratio of the five independant
polynomials is invariant under all the four generating involutions.  In
other words ${\bf C}{\bf P}_9$ is foliated by five dimensional
algebraic
varieties invariant under $\Gamma_3$.

To have some flavour of the possible (integrable ?) algebraic varieties
invariant under $\Gamma_3$, we study its orbits~\cite{bmv1,bmv1b}.  We
start
with the study of the subgroup generated by some infinite order element
namely $IJ$.  This element gives a special role to axis 1.  The
transformation $IJ$ {\em does preserve the symmetry under the exchange
of 2
and 3}.  If the initial point is symmetric under  the exchange of 2 and
3,
the orbit under $IJ$  is thus a {\em curve}.  Other starting points
lead to
orbits lying on a {\em two dimensional} variety given by the
intersection of
seven quadrics (see figure~2,3,4).  However, what we are interested in
are
the orbits of the {\em whole} $\Gamma_3$ group.  The size of this group
prevent us from studying exhaustively the full set of group elements of
a
given length even for quite small values of this length.  We have
nevertheless explored the group by a random construction of typical
elements
of increasingly large length~\cite{bmv2b}.  This confirms that we
generically only have the four invariants described previously.

\vfill\eject \addtocounter{page}{3}

\subsubsection{A second model}

We also consider a simple model where $i$, $j$, $k$, $l$, $m$ and~$n$
take
only two values $+1$ and $-1$ and which is also a generalization of the
Baxter eight vertex model.  The Boltzmann weights $w(i,j,k,l,m,n)$ are
given
by:
\begin{equation}
	w(i,j,k,l,m,n) = f(i,j,k)\;
		\delta^{i}_{l} \;\delta^{j}_{m} \;\delta^{k}_{n}
	+g(i,j,k)\;
		\delta^{i}_{-l}\; \delta^{j}_{-m}\; \delta^{k}_{-n}
		\label{srN}
\end{equation}
\begin{equation}
	f(i,j,k)=f(-i,-j,-k) \mbox{ and }
		g(i,j,k)= g(-i,-j,-k)
\label{sr}
\end{equation}
Equations (\ref{sr}) are symmetry conditions reducing the numbers of
homogeneous parameters from 16 to 8.

As for the previous model, there exists an invariant of the action of
the whole group $\Gamma_3$:
\begin{eqnarray}
{f(+,+,+) f(+,-,-) f(-,+,-) f(-,-,+) \over
g(+,+,+) g(+,-,-) g(-,+,-) g(-,-,+)}
\end{eqnarray}

Considering the subgroup of $\Gamma_3$ generated by the infinite order
element $IJ$, one can easily find other invariants, namely
\begin{eqnarray}
{f(+,+,+) f(+,-,-) \over
g(+,+,+) g(+,-,-)} \\
\noalign{\noindent and}
	{f(+,+,+)^2 +  f(+,-,-)^2 -g(+,+,+)^2 - g(+,-,-)^2 \over
g(+,+,+) g(+,-,-)}
\end{eqnarray}
For this model~\cite{prl}, the trajectories under
$IJ$ are {\em curves} in  ${\bf C}{\bf P}_7$.

\section{Conclusion}

An important problem in  statistical mechanics and field theory, is the
understanding of the role of the dimension of the lattice on both the
algebraic aspects and the topological aspects.  All this touches
various
fields of mathematics and physics:  algebraic geometry, algebraic
topology,
quantum algebra.  Indeed  the Coxeter groups we use are at the same
time
groups of automorphisms of algebraic varieties, symmetries of quantum
Yang-Baxter equations (and their higher dimensional avatars).  They
also
provide an {\em extension to several complex variables functions of the
notion of the fundamental group} ${\bf\Pi}_1$ of a Riemann surface,
with of
course a much more involved covering structure~\cite{JaMa82,bmv1b}.

We believe moreover that the {\em space of parameters} seen as a
projective
space is the appropriate place to look at, if one wants to substantiate
the
deep topological  notion embodied in the notion of ${\bf
Z}$-invariance~\cite{Ba78} and free the models from the details of the
lattice shape.

Actually, we have exhibited an infinite discrete symmetry group for the
Yang-Baxter equations and their higher dimensional generalization
acting on
this parameter space.  This group is the Coxeter group $A_2^{(1)}$
(semi-direct product of ${\bf Z}\times{\bf Z}$ by some finite group).
We
have shown that this symmetry is responsible for the presence of the
spectral parameter.  In other words, the {\em discrete} symmetry gives
rise
to a {\em continuous} one (see~\cite{bmv2}).  A similar study for the
generalized star-triangle relation of the Interaction aRound a Face
model,
sketched in~\cite{MaGa84}, can be performed rigorously along the same
lines,
leading to the same result.  Also note that the same groups generated
by
involutions appear in the study of semi-classical
$r$-matrices~\cite{Av91}.
An interesting point will be to exhibit the action of our symmetry
group on
the underlying quantum group for the Yang-Baxter
equations~\cite{bmvuni}.

Our symmetry group is a group of automorphisms of the  integrability
varieties. This should give  precious informations on these varieties.
In
particular one should decide if, up to Lie groups factors (which cannot
be
excluded because of the existence of ``gauge'' symmetries, weak graph
duality~\cite{GaHi75}, \dots),  these varieties can be anything else
than
abelian varieties, or even product of curves: can they be for example
$K_3$
surfaces, are they homological obstructions to the occurence of
anything but
curves~?

For three-dimensional vertex models, the symmetry group, though
generalizing
very naturally the previous group (generated by four involutions with
similar relations) is drastically different:  it is so
``large''\footnote{
One should keep in mind that very ``large'' sets of rational
transformations
may preserve algebraic curve of genus zero or one. Just think of the
transformations on the circle generated by $\{ \theta \rightarrow
\theta +
\lambda, \; \theta \rightarrow 2 \theta, \; \theta \rightarrow 3 \theta
\}$~\cite{proc3}} that the chances are quite small that it leaves
enough
room for any invariant integrability varieties.  It is not useless to
recall
the unique non-trivial known solution of the tetrahedron equations
(Zamolodchikov's solution)~\cite{Za81,Ba83,Ba86}.  For this model the
three
axes are not on the same footing, so that we do not have a ``true''
three
dimensional symmetry for the model (two-dimensional checkerboard models
coupled together).  Is there still any hope for a three-dimensional
exactly
solvable model with {\em genuine three-dimensional symmetry\/}?  We
think
that the group of symmetries we have described gives the best line of
attack
to this problem.  We will show that $\Gamma_3$ and even more ${\cal
A}ut_3$
are generically too ``large" to allow any non-trivial solution of the
tetrahedron equations with genuine three dimensional
symmetry~\cite{bmvnogo}.


\end{document}